\documentclass[twocolumn,showpacs,amsmath,amssymb,prl,floatfix]{revtex4-1}

\usepackage{graphicx}
\usepackage{float}
\usepackage{dcolumn}
\usepackage{bm}
\usepackage{color}
\usepackage{accents}

\newlength{\dhatheight}

\newcommand{\doublehat}[1]{%
    \settoheight{\dhatheight}{\ensuremath{\hat{#1}}}%
    \addtolength{\dhatheight}{-0.20ex}%
    \hat{\vphantom{\rule{1pt}{\dhatheight}}%
    \smash{\hat{#1}}}}

\begin{document}


\title[Dynamic nuclear polarization as kinetically constrained diffusion]{Dynamic nuclear polarization as kinetically constrained diffusion}

\author{A. Karabanov}
\author{D. Wisniewski}
\affiliation{
Sir Peter Mansfield Magnetic Resonance Centre, School of Physics and Astronomy, University of Nottingham
}%
\author{I. Lesanovsky}
\author{W. K{\"o}ckenberger}%
\email{walter.kockenberger@nottingham.ac.uk}
\affiliation{
Sir Peter Mansfield Magnetic Resonance Centre, School of Physics and Astronomy, University of Nottingham
}%

\date{\today}

\begin{abstract}
Dynamic nuclear polarization (DNP) is a promising strategy for generating a significantly increased non-thermal spin polarization in nuclear magnetic resonance (NMR) applications thereby circumventing the need for strong magnetic fields. Although much explored in recent experiments, a detailed theoretical understanding of the precise mechanism behind DNP is so far lacking. We address this issue by theoretically investigating solid effect DNP in a system where a single electron is coupled to an ensemble of interacting nuclei and which can be microscopically modelled by a quantum master equation. By deriving effective equations of motion that govern the polarization dynamics we show analytically that DNP can be understood as kinetically constrained spin diffusion. On the one hand this approach provides analytical insights into the mechanism and timescales underlying DNP. On the other hand it permits the numerical study of large ensembles which are typically intractable from the perspective of a quantum master equation. This paves the way for a detailed exploration of DNP dynamics which might form the basis for future NMR applications.
\end{abstract}

\pacs{Valid PACS appear here}
\maketitle

The sensitivity of imaging and spectroscopy techniques based on nuclear magnetic resonance (NMR) depends on the spin polarization that arises from the Zeeman interaction of the nuclei with an externally applied magnetic field. However, even with the highest magnetic fields that can currently be generated by superconductive magnets, the nuclear spin polarization of ${}^1$H nuclei is only in the order of less than ten parts per million at ambient temperature. A promising strategy to overcome the low sensitivity in many NMR applications is the use of dynamic nuclear polarization (DNP). Here the electronic spin polarization --- that is about three orders of magnitude higher than the nuclear spin polarization --- is transferred from paramagnetic centres in solid state materials to the surrounding nuclear spin ensemble by irradiating with microwave fields at an appropriate frequency. DNP was already proposed and demonstrated in the early days of nuclear magnetic resonance \cite{overh, slich, jeff}, but it has recently attracted increased attention since progress in hardware technology, mainly in the form of stable high frequency and high power microwave sources, has made it possible to substantially enhance the NMR signal in several applications \cite{prigri, atskoc} including the detection of nuclei with small gyromagnetic constants \cite{maly}, of nuclei located on the surface of structured materials \cite{ross12} and of nuclei in protein micro-crystallites \cite{debe10} by magic angle spinning solid state NMR spectroscopy. Furthermore, DNP carried out at cryogenic temperatures combined with a fast dissolution step to rapidly bring the sample to ambient temperature, was successfully used to generate highly polarised ${}^{13}$C-labelled compounds that could be detected in \emph{in vivo} Magnetic Resonance Imaging applications after their injection \cite{arde03, golm03}. A full understanding of the spin dynamics during DNP is essential for the interpretation of the experimental data and ultimately for optimising the experimental protocols. In principle this problem can be addressed by numerically solving the full quantum master equation \cite{hovav10, v-10}. Owing to its complexity, however, this approach is limited to small systems \cite{us-11,us-12}.

\begin{figure}[ht]
\includegraphics[width=0.46 \textwidth]{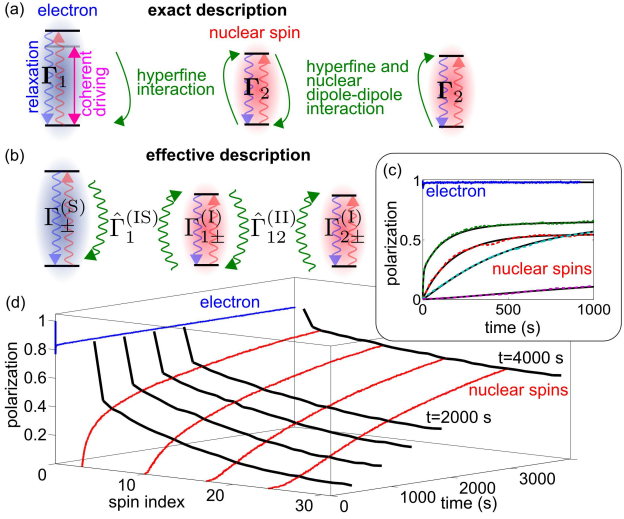}
\caption{(a) Sketch of the solid effect DNP setup. The electron is coherently driven by the applied microwave field and coherent excitation-exchange between the spins is mediated through dipole-dipole couplings. Furthermore the electron and nuclei are subject to dissipation leading to relaxation (see text for details). (b) The effective dynamics of the polarization is described by a classical master equation and thus permits the simulations of large systems. (c) Comparison between the exact (solid black lines) and effective (colored lines) relaxation dynamics of a system consisting of one electron and four nuclei. (d) Polarization dynamics in a linear spin chain with one electron and 30 nuclear spins. The polarization versus spin index (0 denotes the electron position) is shown for different times and the polarization versus time is shown for the electron (blue line) and four selected nuclear spins (red lines).}
\label{fig1}
\end{figure}
In this work we show how to overcome this limitation and to extend the analysis of the spin dynamics to large ensembles containing more than a thousand coupled spins. By using an adiabatic elimination procedure we derive an effective equation for the evolution of the spin polarization. This equation is classical in the sense that it describes transitions between Zeeman basis states and therefore the computational complexity of simulating the spin dynamics is drastically reduced. Moreover, the analytical form of the effective equations of motion permits a transparent interpretation of the physics underlying DNP in terms of operator-valued, i.e kinetically constrained, diffusion rates.

Solid effect DNP (SE DNP) is observed in systems that consist of interacting unpaired electrons $S$ and nuclei $I$ [see Fig. \ref{fig1}(a)] in an external magnetic field with Lamor frequencies $\omega_{\rm S}$ and $\omega_{\rm I}$. Polarization buildup is mediated by the electrons that are irradiated at an off-resonance frequency $\omega_0$ which is in resonance with either the so-called double or zero quantum transition $\omega_{\rm S} \pm \omega_{\rm I}$   \cite{jeff57, abpro58, schje65, jeff63, borgab59, abborg64, v-10}. For SE DNP to be effective, the electrons need to have a resonance linewidth that is smaller than the nuclear Larmor frequency. A suitable minimal model for SE DNP consists of an ensemble of $n$ nuclei coupled to a single unpaired electron through the hyperfine interaction. The spin dynamics of SE DNP can be described by the quantum master equation  $d{\hat\rho}/dt = \bf{L} \hat{\rho} $
with the Liouvillian ${\bf{L}} = -i\doublehat{H}\,\, +{\bf{\Gamma}}$. It describes the coherent evolution of the spin system under the action of the Hamiltonian commutation superoperator $\doublehat H\,\,$ and the effect of the dissipative processes (decoherence) represented by the relaxation superoperator $\bf \Gamma $. Note, that throughout a bold symbol is used to represent superoperators and a doublehat for a commutation superoperator, i.e. $\doublehat{O}\equiv [ \hat O, \cdot ]$.

In the following we discuss the structure of the quantum master equation in detail. The Hamiltonian $\hat H$ of the system consists of four terms: $\hat H = \hat H_{\rm Z} + \hat H_0 +\hat  H_+ + \hat H_-$, where $\hat H_{\rm Z} = \omega_{\rm I} (\hat S_z + \sum_k \hat I_{kz} )$ is the Zeeman term. The term
\begin{eqnarray*}
\hat H_0 = \lambda  \hat S_z + \sum_k A_k \hat I_{kz} \hat S_z + \sum_{k<j}d_{kj}\left(3 \hat I_{kz}\hat I_{jz} - \hat I_k \cdot  \hat I_j\right)
\end{eqnarray*}
describes the hyperfine and nuclear dipole interaction. It depends on the secular strengths of the electron-nuclear hyperfine interaction, parameterized by $A_k$ and furthermore on the inter-nuclear dipolar coupling coefficients $d_{jk}$ \cite{abra}. There is an additional dependence on the so-called offset parameter $\lambda = \omega_{\rm S}-\omega_{\rm I} -\omega_0$ which represents a resonance condition between the microwave frequency $\omega_0$ and that of the double quantum transition $\omega_{\rm S}-\omega_{\rm I}$. The remaining terms of the system Hamiltonian are
\begin{eqnarray*}
\hat H_\pm = \frac{\omega_1}{2} \hat S_\pm + \frac{1}{2}\sum_k B_{k\pm} \hat I_{k\pm} \hat S_z  ,
\end{eqnarray*}
which arise from the pseudosecular hyperfine interaction and the microwave irradiation. They are parameterized by the pseudosecular coupling strength $B_{k\pm}$ between the electron and the nuclei and the microwave amplitude $\omega_1$. Note, that the interaction parameters  $A_k$, $B_{k\pm}$ and $d_{kj}$ depend on the geometry of the spin system \cite{abra}.

The relaxation superoperator is composed of two terms, ${\bf\Gamma} = {\bf\Gamma}_1 + {\bf\Gamma}_2$. The first one is given by
\begin{eqnarray*}
{\bf\Gamma}_1 &=& \frac{R_1^{(\mathrm{S})}}{2}[{\bf D}(\hat S_+)+{ \bf D}( \hat S_-)]+{\bf\Gamma}_{\mathrm{th}} \\
&+& \frac{R_1^{(\mathrm{I})}}{2}\sum_k[{ \bf D}(\hat I_{k+})+ {\bf D}(\hat I_{k-})]
\end{eqnarray*}
where we have defined the dissipator ${\bf D}(\hat X) \hat\rho \equiv \hat X \hat \rho \hat X^{\dagger} -  \{ \hat \rho, \hat X^{\dagger} \hat X  \}/2$. ${\bf\Gamma}_1$ describes dissipative processes or longitudinal relaxation due to interaction of the spins with the environment \cite{us-14}.  The rates for longitudinal relaxation are $R_1^{(\rm S)}$ and $R_1^{(\rm I)}$ for the electron and the nuclei, respectively. The superoperator ${\bf\Gamma}_{\mathrm{th}}=P_0\,R_1^{(\mathrm{S})}[ {\bf D}(S_-)-{\bf D}(S_+)]/2$ acts as a correction term and ensures that the electron relaxes back to its thermal equilibrium state which---at typical temperatures of $T > 0.1$K--- is up to a scaling factor approximated by $\rho_{\rm {th}} = 1-2P_0 \hat S_z$. Here $P_0 = \tanh \left (\hbar \omega_{\rm S}/2k_{\rm B} T \right )$ is the thermal electronic polarization \cite{abra}. The second term,
\begin{eqnarray*}
{\bf\Gamma}_2 =2R_2^{(\mathrm{S})} {\bf D}(S_z)+2\sum_kR_2^{(\mathrm{I})} {\bf D}(I_{kz})
\end{eqnarray*}
represents the loss of coherences or transverse relaxation of the spin ensemble, which depends on the rates $ R_2^{(\rm S)}$ and $ R_2^{(\rm I)} $ for the electron and the nuclear spins, respectively.

The full quantum master equation describes the dynamics of the spin system in the Liouville space ${\cal L}$ of dimension $2^{2N}$ where $N = n+1$. However, all information that is relevant for characterizing SE DNP is typically contained in the Zeeman subspace ${\cal L}_{\rm Z}$ \cite{smith} which is spanned by the operators $\{\hat1,\,\hat I_{kz},\,\hat I_{kz} \hat I_{k'z},\, \hat I_{kz}\hat I_{k'z}\hat I_{k''z},\,\ldots\}$ and has a dimension $2^N$. For example the bulk nuclear polarization is given by the expectation value of $\sum_n \hat I_{kz}$. We have obtained a closed equation for the dynamics in the Zeeman subspace by means of a quite lengthy adiabatic approximation procedure which is detailed in the supplemental material. This resulting effective master equation is of the form $d{\hat \rho}_{\rm Z}/dt = {\bf L}_{\rm Z} \hat \rho_{\rm Z}$. It has not only a substantially reduced dimension but in addition it is of entirely classical nature --- although we employ in the following a notation that is also used in the quantum case. As shown in the supplemental material the dynamics can be expressed in terms of jump operators [see Fig. \ref{fig1}(b)], which connect classical spin configurations, and their corresponding effective jump rates, i.e. the equation is of Lindblad form \cite{lind76}. Similar approaches have been employed in the study of cold atomic systems \cite{berni14, polet13, marcu14, lesan13}. When formulated in this way the problem lends itself to the use of classical kinetic Monte Carlo (kMC) algorithms which permits the simulation of large systems. To check the validity of this approximate approach we have simulated small systems with the exact quantum Master equation and kMC. Example results are shown in Fig. \ref{fig1}(c) as well as in the supplemental material and show excellent agreement between the two methods.

The effective dynamics in the Zeeman subspace is governed by a set of single-spin and two-spin dissipators: ${\bf L}_{\rm Z} = {\bf L}_{\mathrm{single-spin}} + {\bf L}_{\mathrm{two-spin}}$. The single-spin terms are given by
\begin{eqnarray}
{\bf L}_{\mathrm{single-spin}} &=& \Gamma^{(\rm{S})}_+{\bf D}(\hat S_+)+\Gamma^{(\rm{S})}_- {\bf D}(\hat S_-)+  \nonumber\\
& &+\sum_k\left[\Gamma^{(\rm{I})}_{k+} {\bf D}(\hat I_{k+})+\Gamma^{(\rm{I})}_{k-} {\bf D}(\hat I_{k-})\right] \label{L1}.
\end{eqnarray}
and depend on the constant rates
\begin{eqnarray}
\Gamma^{(\mathrm{S})}_\pm &=& \frac{1\mp P_0}{2}R_1^{(\rm S)} + \frac{\omega_1^2}{2\omega_{\rm I}^2}R_2^{(\rm S)}, \label{g1}\\
\Gamma^{(\rm I)}_{k\pm} &=& \frac{R_{1}^{(\rm I)}}{2} + \frac{B_{k-}B_{k+}}{8\omega_{\rm I}^2}R_{2}^{(\rm I)}. \label{g2}
\end{eqnarray}
Here $\Gamma^{(\mathrm{S})}_\pm$ are the rates at which the electron spin flips 'up' ($+$) or 'down' ($-$). They depend on the longitudinal relaxation rate $R_1^{(\rm S)}$ of the electron with a weighting pre-factor which ensures that in the absence of any perturbation by the microwave field the electronic steady-states polarization is identical to the thermal electronic polarization $P_0$. The second term is a consequence of the applied  microwave field. Eq. (\ref{g2}) describe the rates for a nuclear spin $k$ to flip 'up' or 'down'. Since we have assumed in the derivation of these equations that the thermal nuclear polarization is negligibly small, these two rates coincide and correspond to half the longitudinal nuclear relaxation rate $R_{1}^{(\rm I)}$. The additional term appearing in Eq. (\ref{g2}) arises from the pseudosecular interaction with the electron which leads to a tilt of the nuclear effective field axis.
\begin{figure*}
\includegraphics[width=\textwidth]{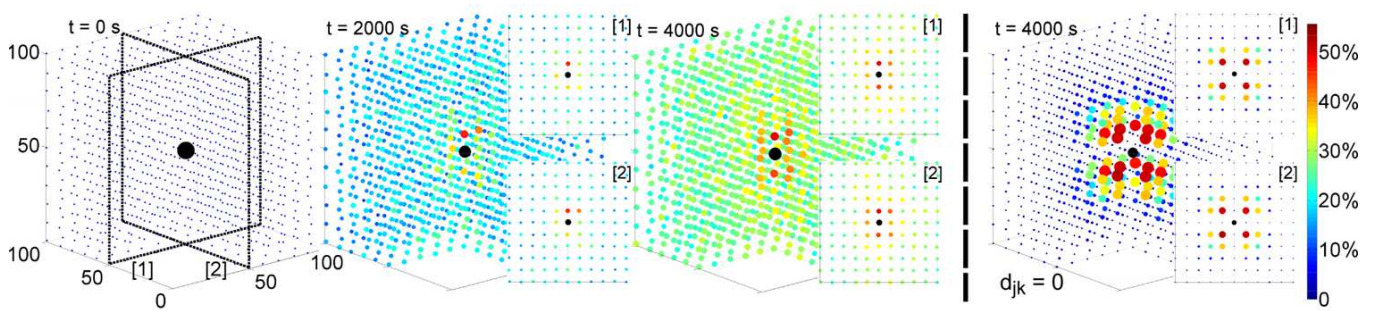}
\caption{\label{fig2}Simulation of the polarization dynamics of 1330  nuclear spins (${}^{13}C$) arranged on a regular grid in a cube ($11 \times 11 \times 11$) around one central electron (black). The diameter and the colour of the spheres indicate the expectation value of the nuclear spin polarization at the given position. The simulation demonstrates that polarization buildup is fastest for the nuclei that are located close to the electron at an angle of $\pi/4$ with respect to direction of the applied static magnetic field. In order to show the contribution of spin diffusion to the distribution of polarization within the spin ensemble the dipolar coupling constants were set to zero in the rightmost panel. This prevents any transport of polarization through spin flip-flop transitions between pairs of nuclear spins. Note, that the front quarter of the ${}^{13}C$ spin ensemble is not shown for better visibility of the polarization buildup near the electron. }
\end{figure*}

The two-spin dissipators are given as
\begin{equation}
{\bf L}_{\mathrm{two-spin}} =  \sum_k \hat\Gamma^{(\rm{IS})}_k {\bf D}( \hat Y_k)  + \sum_{k<j} \hat\Gamma^{(\rm{II})}_{kj} {\bf D}( \hat X_{kj}),
\label{L2}
\end{equation}
with
$
 \hat Y_k=\hat I_{k+} \hat S_-+\hat I_{k-} \hat S_+$ and $
 \hat X_{kj}=\hat I_{k+}\hat I_{j-}+\hat I_{k-} \hat I_{j+}\;.
$
They clearly describe a diffusive (flip-flop) dynamics between the electron and the nuclei as well as among the nuclei. Interestingly the diffusion rates $\Gamma^{(\rm{IS})}_k$ and $\hat\Gamma^{(\rm{II})}_{kj}$ are not constant, but operator-valued. Hence the diffusion rate of the spin-polarization depends on the state of the entire spin system. In the context of the study of glassy relaxation \cite{Ritort,Garrahan} such operator-valued rates are often referred to as \emph{kinetic constraint}. Explicitly, they read
\begin{eqnarray}
\hat \Gamma^{(\rm{IS})}_k &=&\frac{\omega_1^2}{8\omega_I^2} \frac{\left|B_{k-}B_{k+}\right|}{ R_2^{(\rm S)} + R_{2}^{(\rm I)}  } \left (\hat 1+\hat D_k^2 \right )^{-1} \nonumber ,\\
\hat D_k&=&\frac{\lambda \hat 1 +
\sum_{s\not=k}A_s \hat I_{sz}}{R_2^{(\rm S)} + R_{2}^{(\rm I)}}, \label{g3}
\end{eqnarray}
\begin{eqnarray}
\hat \Gamma^{(\rm{II})}_{kj} = \frac{d_{kj}^2}{4 R_{2}^{(\rm I)} } \left ( \hat 1+\hat C_{kj}^2 \right )^{-1},\,
\hat C_{kj}=\frac{(A_k-A_j)\hat S_z}{ 2 R_{2}^{(\rm I)}}. \label{g4}
\end{eqnarray}
A brief discussion of these expressions provides insight into the spin dynamics on which SE DNP is based on. Eq. (\ref{g3}) describes the rate of flip-flop transitions or jumps between a nuclear spin $k$ and the unpaired electron. Hence $\hat \Gamma^{(\rm{IS})}_k$ is of fundamental importance for the SE DNP effect. This rate depends on both the strength of the microwave field $\omega_1$ and the pseudosecular interaction strength $B_{k\pm}$. It is inversely proportional to the square of the nuclear Larmor frequency $\omega_\mathrm{I}$ and also inversely proportional to the sum of the transverse relaxation rates of the electron, $R_2^{(\rm S)}$, and the nuclear spin, $R_2^{(\rm I)}$. This indicates that although this process is represented by the flip-flop jump operator, the underlying quantum mechanical process is mediated by coherences which in fact decay with these relaxation rates.

Most importantly, the rate for polarization transfer between the electron and the $k$-th nucleus is controlled by the operator $\hat D_k$ whose (eigen)value depend on the polarization level of the nuclear spin ensemble. For negligible nuclear polarization in an ensemble of nuclear spins the sum $\sum_{s\not=k}A_s \hat I_{sz}$ evaluates to a value close to zero as there is an equal probability for the nuclei to be in their spin 'up' or 'down' state. However, with increasing level of nuclear polarization the magnitude of this sum grows which decreases the effective rate $\hat\Gamma_k^{(\rm IS)}$. Note, that the effect of the off-resonance parameter $\lambda$ is negligible as long as the microwave frequency $\omega_0$ is matched to the double quantum transition at $\omega_{\rm S} - \omega_{\rm I}$.

The transport of polarization within the dipolar-coupled network of the nuclear spins takes places via spin diffusion which is governed by the operator-valued rate given in Eq. (\ref{g4}). The rate of polarization transport among two nuclei is proportional to the square of their nuclear dipolar coupling constant $d_{kj}$, and it is indirectly proportional to the nuclear transverse relaxation rate $R_2^{(\rm{I})}$. Interestingly, there is also a dependence on the difference of the secular hyperfine interaction strength $A_k$ of the two nuclei. In the case in which this difference is significant, the internuclear polarization transfer between the spins can become strongly quenched provided the electronic polarization $\hat{S}_z$ is non-zero.

As the effective dynamics in the Zeeman subspace can be treated efficiently numerically SE DNP can be studied in relatively large samples. In Fig. \ref{fig1}(d) we already showed a simple example for 31 spins in one dimension while Fig. \ref{fig2} now shows the progressive buildup of spin polarization within a three-dimensional sample of 1330 ${}^{13}$C spins arranged on a regular cubic lattice with one electron spin at the centre. For the latter the distance between the nuclei is 10 $\pm \;5 \% \; \AA $ and the field strength of the static external magnetic $B_0$ field is 3.4 T. The applied microwave field has a frequency of $\omega_0 = \omega_{\rm S} - \omega_{\rm I} = 34.5$ MHz and a strength $\omega_1=$100 kHz. The relaxation time constants are $R_1^{(\mathrm{S})} = 1\: s^{-1}, R_2^{(\mathrm{S})}= 10^5 \: s^{-1}, R_1^{(\mathrm{I})} = 1.4 \times 10^{-4} \:s^{-1}$ and $R_2^{(\mathrm{I})} = 10^4 \: s^{-1}$.

Let us finally investigate the role the nuclear dipolar interaction plays in the distribution of polarization between the nuclear spins. This question has been addressed e.g. by Hovav \emph{et al.} in Ref. \cite{hovav10} who conducted simulations of small chains (up to 9 nuclei) within the framework of the full quantum master equation. Their results show that there is only a weak dependence of the polarization build up rates on the nuclear dipolar couplings between nuclei further away from the electron (nuclei belonging to the bulk) suggesting that there is a direct transfer of polarization from the electron to all bulk nuclei through mixing of the multi-nuclear product states. Our data shown in Fig. \ref{fig2} clearly suggests that the nuclear dipolar couplings are important. To further investigate their role in polarization transport we have simulated the polarization dynamics of a linear spin chain consisting of 30 spins as shown in Fig. \ref{fig1}(d) [see supplemental material for further details]. In contrast to the analysis in Ref. \cite{hovav10} our findings show that a decrease of the nuclear dipolar interaction constants of the bulk nuclei by a factor of two alters the polarization buildup dramatically. This is consistent with a diffusive polarization transport. The discrepancy is thus most likely caused by boundary effects, which can be severe for short chains.
Moreover, further discrepancies might arise due to the way in which relaxation was introduced in the reference frame in Ref. \cite{hovav10} as was already previously discussed in Ref. \cite{us-14}. 

Let us now establish a more quantitative connection between SE DNP and the diffusion of polarization. For a linear chain of $n$ nuclear spins, as shown in Fig. \ref{fig1}, we can derive an average diffusion constant:
\begin{equation}
D_{\rm av} = \frac{1}{n-1} \sum ^{n-1}_{k=1} D_{k}, \, D_k = \frac{4 R_2^{(\rm I)} d_k^2 a_k^2}{\left (4 R_2^{(\rm I)}\right )^2 +  \left(A_k - A_{k+1} \right)^2}
\label{dif}
\end{equation}
Here $a_k$ is the distance between the nuclear spins $k$ and $k+1$, $d_{k}$ is the dipolar coupling constant between the two spins and $A_k$ is the strength of the secular hyperfine interaction of spin $k$. The constant $D_k$ is the rate of diffusive polarization transport among neighboring nuclei. It is quenched when the difference between the secular hyperfine interactions of the two spins, $\vert A_k  - A_{k+1} \vert $, is large and when the nuclear transverse relaxation rate $R_2^{(\rm I)}$ is fast. Comparing the polarization dynamics calculated via kMC with a simple diffusion model, $ \partial p/ \partial t =  D_{\rm av} \frac{\partial^2 p}{\partial x^2}$, where $p \equiv p(x,t)$ is the polarization field, indeed shows good agreement with data as the one presented in Fig. \ref{fig1}(d). Further details and a discussion of the relevant boundary conditions is provided in the supplemental material. We conclude that polarization transport in a SE DNP experiment is mediated by nuclear spin diffusion. The presence of the electron will hinder --- or constrain --- this process for nuclei close to the electron since for these nuclei the difference between the strengths of the secular hyperfine interactions is significant \cite{afe92, hovav10}. 

We have derived effective equations of motions which show that the mechanism underlying SE DNP is kinetically constrained diffusion. The derived set of equations enable the study polarization buildup in relatively large spin systems which has been demonstrated by investigating DNP in a system with 1330 ${}^{13}$C nuclei and a single electron. Finally, we have highlighted the importance of nuclear dipolar couplings in SE DNP and derived an effective rate for polarization diffusion in a one-dimensional chain. We believe that this study represents an important step in the quantitative understanding of SE DNP for practical applications in NMR. In the future we will investigate whether the method presented here is also applicable in more complex settings, which involve e.g. more than a single electron.

\begin{acknowledgments}
We are grateful to Prof Shimon Vega, Weizman Institute, Rehovot, for interesting discussions related to DNP. The research leading to these results has received funding from the European Research Council under the European Union's Seventh Framework Programme (FP/2007-2013) / ERC Grant Agreement No. 335266 (ESCQUMA),the EU-FET grants HAIRS 612862 and RYSQ 640378 and EPSRC Grant EP/I027254/1.
\end{acknowledgments}

\newpage

\onecolumngrid
\section{Supplementary Material} 

\subsection{Adiabatic Elimination}

For any linear system with constant coefficients
\begin{equation}
\dot{\hat\sigma}={\bf L}\hat\sigma
\label{i}
\end{equation}
defined in some vector space ${\cal H}$, the following technique is valid.

Suppose we need to know dynamics in a subspace ${\cal H}_1\subset{\cal H}$ and initially also $\hat\sigma(0)\equiv\hat\sigma_1(0)\in{\cal H}_1$. By using the space decomposition
$$
{\cal H}={\cal H}_1+{\cal H}_2
$$
and introducing the projections
$$
\hat\sigma_k={\bf\pi_k}\hat\sigma,\quad
{\bf L_{kj}}={\bf\pi_k}{\bf L}{\bf\pi_j},\quad
k,j=1,2,
$$
we can write
$$
\dot{\hat\sigma}_1={\bf L_{11}}\hat\sigma_1+{\bf L_{12}}\hat\sigma_2,\quad
\dot{\hat\sigma}_2={\bf L_{22}}\hat\sigma_2+{\bf L_{21}}\hat\sigma_1.
$$
Given $\hat\sigma_1$, the second equation is resolved as
$$
\hat\sigma_2(t)=e^{{\bf L_{22}}t}\int_0^te^{-{\bf L_{22}}\tau}{\bf L_{21}}\hat\sigma_1(\tau)\,d\tau.
$$
Substituting this into the first equation, we obtain the equation closed in ${\cal H}_1$
\begin{equation}
\dot{\hat\sigma}_1(t)={\bf L_{11}}\hat\sigma_1(t)+\int_0^t{\bf K}(t-\tau)\hat\sigma_1(\tau)\,d\tau,
\label{p}
\end{equation}
where
\begin{equation}
{\bf K}(T)={\bf L_{12}}e^{{\bf L_{22}}T}{\bf L_{21}}.
\label{fK}
\end{equation}

In terms of the Laplace transforms (defined for $\zeta>0$)
$$
\hat l_1(\zeta)=\int_0^{+\infty}e^{-\zeta t}\hat\sigma_1(t)\,dt,\quad
{\bf l_K}(\zeta)=\int_0^{+\infty}e^{-\zeta t}{\bf K}(t)\,dt,
$$
(\ref{p}) is resolved as
\begin{equation}
\hat l_1(\zeta)=({\bf P}(\zeta))^{-1}\hat\sigma_1(0),
\label{lp}
\end{equation}
where
\begin{equation}
{\bf P}(\zeta)=\zeta{\bf 1}-{\bf L_{11}}-{\bf l_K}(\zeta).
\label{PQ}
\end{equation}
The initial value $\hat\sigma_1(0)$ and the Laplace transform of the kernel ${\bf l_K}$ uniquely define the Laplace transform of the solution. The solution itself is then uniquely found by inversion of its Laplace transform.

It follows from (\ref{p}) that the solution at time $t$ depends on its history in the time interval $[0,t]$. The integral kernel ${\bf K}$ is often called {\it memory function}. Besides the memory effects, ${\bf K}$ describes how dynamics in the complementary subspace ${\cal H}_2$ affect the dynamics in ${\cal H}_1$.

If all eigenvalues $\zeta_k$ of the superoperator ${\bf L_{22}}$ have negative real parts then any solution to (\ref{p}) tends to the steady-state solution
$$
\hat\sigma_1(t)\to\hat\sigma_{1*},\quad
t\to+\infty,
$$
satisfying
$$
({\mathbf L_{11}}+{\bf M})\hat\sigma_{1*}=0
$$
with
$$
{\mathbf M}=\int_0^{+\infty}{\bf K}(t)\,dt=-{\bf L_{12}}{\bf L_{22}}^{-1}{\bf L_{21}}.
$$

On the other hand, in terms of (\ref{fK}), (\ref{PQ}) and the Laplace transform,
$$
{\bf K}(t)=\sum e^{\zeta_kt}{\bf K_k},\quad
{\bf l_K}(\zeta)=\sum\frac{\bf K_k}{\zeta-\zeta_k}.
$$
Due to the formulas
$$
{\bf l_K}(0)=-\sum\frac{\bf K_k}{\zeta_k}={\bf M},\quad
\frac{d{\bf l_K}}{d\zeta}(0)=-\sum\frac{\bf K_k}{\zeta_k^2},
$$
we have that under the conditions
$$
\vert\zeta\vert\ll\zeta_-,\quad
\Vert{\bf L_{12}}\Vert\cdot\Vert{\bf L_{21}}\Vert\ll\zeta_-^2,\quad
\zeta_-=\min\,\vert\zeta_k\vert
$$
we obtain the approximation
$$
{\bf l_K}(\zeta)={\bf l_K}(0)+\frac{d{\bf l_K}}{d\zeta}(0)\zeta+\ldots\sim{\bf M}
$$
where the second term is negligibly small. Since the small values of $\zeta$ are responsible for the long-term slow dynamics, it means that the steady-state part $\bf M$ of the integral term well approximates the informative long-term asymptotic of the dynamics.   

Thus, under the conditions
\begin{equation}
\Vert{\bf L_{12}}\Vert\cdot\Vert{\bf L_{21}}\Vert\ll\zeta_-^2,\quad
\Vert{\bf L_{11}}+{\bf M}\Vert\ll\zeta_-,
\label{ltc}
\end{equation}
we have the asymptotics in (\ref{PQ})
$$
{\bf P}\sim\zeta{\bf 1}-{\bf L_{11}}-{\bf M}.
$$
Returning to the Laplace transform (\ref{lp}), this means that solutions to equation (\ref{p}) are well approximated by solutions to the equation
\begin{equation}
\dot{\hat\sigma}_1(t)=({\bf L_{11}}+{\bf M})\hat\sigma_1(t).
\label{lta}
\end{equation}
Equation (\ref{lta}) does not contain any memory function. Its operator ${\bf L_{11}}+{\bf M}$ is time-independent. Due to (\ref{ltc}), fast dynamics in the complementary subspace ${\cal H}_2$ make the solution in ${\cal H}_1$ tend to the steady-state, rapidly ``forgetting'' its previous history.

We call equation (\ref{lta}) an {\it adiabatic approximation} for equation (\ref{p}) and say that the complementary subspace ${\cal H}_2$ is {\it adiabatically eliminated}. The term {\it adiabatic} means that we first separate fast motions in ${\cal H}_2$ from slow motions in ${\cal H}_1$ and then eliminate all information about the fast motions, not influencing the slow dynamics. Here the equation preserves its initial form of the linear system with constant coefficients. The technique is easily generalised to inhomogeneous systems and initial conditions outside the informative subspace ${\cal H}_1$.

Such procedures are commonly used (among other fields) in condensed matter theory, quantum optics (where the term {\it adiabatic elimination} initially comes from), quantum statistics and quantum information, in cases where either external driving or an interaction with an environment are involved in such way that a non-resonant part of the dynamics becomes non-informative. A similar effect, well-known in the NMR context, is achieved by proceeding to a rotating frame and averaging out secular terms. Another example is eliminating high spin correlation orders in multi-spin dynamics where the role of the fast motion is played by spin relaxation. See in this context, for example, \cite{zg, us-12,us-11}. If the dimension of the informative subspace ${\cal H}_1$ is much smaller than the dimension of the total space ${\cal H}$, this gives a significant reduction of the initial problem.

\subsection{Elimination of non-zero quantum coherences}

The Liouville state space can be decomposed as
$$
{\cal L}=\sum{\cal L}_q,
$$
where ${\cal L}_q$ is the subspace of $q$-quantum coherences,
\begin{equation}
{\doublehat H}_{\,\,Z}\hat\rho=q\omega_I\hat\rho,\quad
\hat\rho\in{\cal L}_q,\quad
q=0,\,\pm 1,\,\ldots,\,\pm(n+1).
\label{Qq}
\end{equation}
Within the method described in the previous section, let the subspace ${\cal H}_1$ be the subspace of zero-quantum coherences ${\cal L}_0$ and ${\cal H}_2$ the complementary subspace,
$$
{\cal H}_1={\cal L}_0,\quad
{\cal H}_2=\sum_{q\not=0}{\cal L}_q.
$$

Starting from the thermal equilibrium, we have
$$
\hat\sigma_1(0)=1-2P_0\hat S_z.
$$
The relations
$$
(i{\doublehat H}_{\,\,0}+{\bf\Gamma}){\cal L}_q\subset{\cal L}_q,\quad
{\doublehat H}_{\,\,\pm}{\cal L}_q\subset{\cal L}_{q\pm 1}
$$
along with (\ref{Qq}) give
\begin{equation}
\begin{array}{c}
{\bf L_{11}}\hat\sigma_1=-(i{\doublehat H}_{\,\,0}+{\bf\Gamma})\hat\sigma_1,\quad
{\bf L_{21}}\hat\sigma_1=-i({\doublehat H}_{\,\,+}+{\doublehat H}_{\,\,-})\hat\sigma_1,\\
{\bf L_{12}}{\cal L}_{\pm 1}=-i{\doublehat H}_{\,\,\mp}{\cal L}_{\pm 1},\quad
{\bf L_{12}}{\cal L}_q=0,\quad
\vert q\vert>1,\\
{\bf L_{22}}{\cal L}_{\pm 1}=-(\pm i\omega_I{\bf 1}+i{\doublehat H}_{\,\,0}+i{\doublehat H}_{\,\,\pm}+{\bf\Gamma}){\cal L}_{\pm 1},\\
{\bf L_{22}}{\cal L}_q=-(iq\omega_I{\bf 1}+i{\doublehat H}_{\,\,0}+i{\doublehat H}_{\,\,+}+i{\doublehat H}_{\,\,-}+{\bf\Gamma})
{\cal L}_q,\quad
\vert q\vert>1.
\end{array}
\label{r}
\end{equation}

Due to the presence of relaxation, the eigenvalues of the superoperator $\bf L_{22}$ have negative real parts, satisfying the requirement outlined in the previous section. Physically, at high magnetic field,
\begin{equation}
\Vert\omega_I^{-1}{\doublehat H}_{\,\,0,\pm}\Vert,\,\Vert\omega_I^{-1}{\bf\Gamma}\Vert\sim\epsilon\ll 1,
\label{phys}
\end{equation}
so the superoperator $\bf M$ is found as the series
\begin{equation}
{\bf M}=\omega_I^{-1}{\bf M_1}+\omega_I^{-2}{\bf M_2}+\ldots
\label{M}
\end{equation}
converging exponentially fast
$$
{\bf M}-\sum_{k=1}^m\omega_I^{-k}{\bf M_k}\sim\epsilon^{m+1}.
$$
Using (\ref{r}), we can calculate
\begin{equation}
{\bf M_1}=-i[\doublehat H_{\,\,+},\doublehat H_{\,\,-}],\quad
{\bf M_2}=-\doublehat H_{\,\,+}(i\doublehat H_{\,\,0}+{\bf\Gamma})\doublehat H_{\,\,-}-\doublehat H_{\,\,-}
(i\doublehat H_{\,\,0}+{\bf\Gamma})\doublehat H_{\,\,+},\quad
\ldots
\label{M2}
\end{equation}

On the other hand, it follows from (\ref{r})
$$
\zeta_->\vert\omega_I\vert,\quad
\Vert{\bf L_{12}}\Vert\le\max\Vert\doublehat H_{\,\,\pm}\Vert,\quad
\Vert{\bf L_{21}}\Vert\le 2\max\Vert\doublehat H_{\,\,\pm}\Vert.
$$
Hence, because of (\ref{phys}), (\ref{M}), (\ref{M2}), conditions (\ref{ltc}) are satisfied, so we can replace equation (\ref{p}) by its adiabatic approximation
\begin{equation}
\dot{\hat\sigma}={\bf L_0}\hat\sigma
\label{plt}
\end{equation}
with
$$
{\bf L_0}={\bf L_{11}}+{\bf M}={\bf M_0}+\omega_I^{-1}{\bf M_1}+\omega_I^{-2}{\bf M_2},
$$
where
$$
{\bf M_0}=-(i\doublehat H_{\,\,0}+{\bf\Gamma}),
$$
${\bf M_{1,2}}$ are defined in (\ref{M2}), and in (\ref{M}) we restricted to the second order approximation for $\bf M$.

\subsection{Elimination of non-Zeeman spin orders}

In the previous section, we adiabatically proceeded from the initial Liouvillian $\bf L$ to the new Liouvillian $\bf L_0$, well describing the dynamics closed in the subspace ${\cal L}_0$ of zero-quantum coherences.

We decompose now the zero-quantum subspace as
$$
{\cal L}_0={\cal L}_Z+{\cal L}_C,
$$
where
$$
{\cal L}_Z=span\{\hat 1,\,\hat I_{kz},\,\hat I_{kz}\hat I_{k'z},\,\hat I_{kz}\hat I_{k'z}\hat I_{k''z},\,\ldots\}
$$
is the subspace of Zeeman spin orders, and ${\cal L}_C$ is the complementary subspace consisting of non-Zeeman zero-quantum coherences.

The commutation character of the notation $\doublehat O\equiv[\hat O,\cdot]$ implies that the superoperator $\bf M_1$ is a commutation,
$$
{\bf M_1}=-i\doublehat H_{\,\,1},\quad
\hat H_1=[\hat H_+,\hat H_-]=\hat H_{1,0}+\hat H'_1,
$$
$$
\hat H_{1,0}=\frac{\omega_1^2}{2}\hat S_z+\frac{1}{8}\sum B_{k+}B_{k-}\hat I_{kz},\quad
\hat H'_1=-\frac{\omega_1}{4}\sum(B_{k+}\hat I_{k+}\hat S_-+B_{k-}\hat I_{k-}\hat S_+).
$$
We have also
$$
\hat H_0=\hat H_{0,0}+\hat H'_0,\quad
{\bf\Gamma}={\bf\Gamma_0}+{\bf\Gamma'},
$$
$$
\hat H_{0,0}=\lambda\hat S_z+\sum_kA_k\hat I_{kz}\hat S_z,
$$
$$
\hat H'_0=\sum_{j<k}d_{kj}\left[2\hat I_{kz}\hat I_{jz}-\frac{1}{2}\left(\hat I_{k+}\hat I_{j-}+\hat I_{k-}\hat I_{j+}\right)\right],
$$
$$
{\bf\Gamma_0}=R_2^{(S)}\doublehat S_{\,\,z}^{\,\,\,2}+\sum R_{2}^{(I)}\doublehat I_{\,\,kz}^{\,\,\,2},
$$
\begin{equation}
{\bf\Gamma'}=\frac{R_1^{(S)}}{4}\left(\doublehat S_{\,\,+}\doublehat S_{\,\,-}+\doublehat S_{\,\,-}\doublehat S_{\,\,+}\right)+
\frac{1}{4}\sum R_{1}^{(I)}\left(\doublehat I_{\,\,k+}\doublehat I_{\,\,k-}+\doublehat I_{\,\,k-}\doublehat I_{\,\,k+}\right)+{\bf\Gamma_{th}}.
\label{G'}
\end{equation}

The superoperators $\doublehat H_{\,\,0,0},\,\doublehat H_{\,\,1,0},\,{\bf\Gamma_0}$ trivially act on ${\cal L}_Z$. Within the accuracy of $\sim\omega_I^{-1}$,  only the superoperators $\doublehat H_{\,\,0}^{\,\,\,'}$, $\doublehat H_{\,\,1}^{\,\,\,'}$ contribute to ${\bf L_{21}},\,{\bf L_{12}}$. The superoperator
$$
{\bf X}=-i\doublehat H_{\,\,0,0}-{\bf\Gamma_0}
$$
maps ${\cal L}_C$ to itself with eigenvalues $\zeta'_k$ satisfying
$$
Re\,\zeta'_k<0,\quad
\vert\zeta'_k\vert> \min\{2R_{2}^{(I)},\,R_2^{(S)}+R_{2}^{(I)}\}.
$$
Thus, under the conditions
\begin{equation}
\min \Big\{\left (2R_{2}^{(I)} \right)^2,\,\left (R_2^{(S)}+R_{2}^{(I)}\right)^2\Big\} \gg \max\Big\{\frac{\vert d_{kj}^2\vert}{4},\,\frac{\vert\omega_1 B_k\vert^2}{16\vert\omega_I\vert^2},\,\left(R_1^{(S)}\right)^2,\,\left(R_{1}^{(I)}\right)^2\Big\},
\label{aac}
\end{equation}
conditions (\ref{ltc}) are satisfied and the subspace ${\cal L}_C$ is adiabatically eliminated. We come then to the adiabatic approximation
\begin{equation}
\dot{\hat\sigma}={\bf L_Z}\hat\sigma,\quad
{\bf L_Z}={\bf L_{0,11}}+{\bf M'},\quad
{\bf M'}=-{\bf L_{0,12}}{\bf L_{0,22}}^{-1}{\bf L_{0,21}}
\label{aaZ}
\end{equation}
closed in the Zeeman subspace ${\cal L}_Z$. In accordance with the previous notations,
$$
{\bf L_{0,kj}}={\bf\pi'_k}{\bf L_0}{\bf\pi'_j},\quad
k,\,j=1,\,2,
$$
where $\bf\pi'_{1,2}$ are projections onto the subspaces ${\cal L}_{Z,C}$ respectively.

Using conditions (\ref{phys}) , (\ref{aac}), it can be shown that the right-hand side of (\ref{aaZ}) is well approximated as
\begin{equation}
\begin{array}{c}
{\bf L_{0,11}}=-{\bf\Gamma'}-{\bf\Gamma''},\\[3mm]
{\displaystyle{\bf M'}=-\sum_{j<k}\hat C'_{kj}\doublehat X_{\,\,kj}^{\,\,\,'2}-\sum_k\hat D'_k\doublehat Y_{\,\,k}^{\,\,\,'2},}
\end{array}
\label{fa}
\end{equation}
where $\bf\Gamma'$ is given by (\ref{G'}),
$$
{\bf\Gamma''}=\frac{1}{4\omega_I^2}\left[\omega_1^2R_2^{(S)}\left(\doublehat S_{\,\,+}\doublehat S_{\,\,-}+\doublehat S_{\,\,-}
\doublehat S_{\,\,+}\right)+
\frac{R_2^{(I)}}{4}\sum_kB_{k+}B_{k-}\left(\doublehat I_{\,\,k+}\doublehat I_{\,\,k-}+\doublehat I_{\,\,k-}
\doublehat I_{\,\,k+}\right)\right],
$$
and
\begin{equation}
\begin{array}{c}
{\displaystyle\hat X'_{kj}=\frac{d_{kj}}{2}(\hat I_{k+}\hat I_{j-}+\hat I_{k-}\hat I_{j+}),}\\[4mm]
{\displaystyle\hat Y'_k=\frac{\omega_1}{4\omega_I}(B_{k+}\hat I_{k+}\hat S_-+B_{k-}\hat I_{k-}\hat S_+),}\\[4mm]
{\displaystyle\hat C'_{kj}= \frac{2 R_{2}^{(I)}}{\left(2R^{(I)}_2\right)^2 +\hat C^2_{kj}},\quad
\hat D'_k=\frac{R_2^{(S)} + R_2^{(I)}} {\left ( R_2^{(S)} + R_2^{(I)} \right)^2+\hat D^2_k},}\\[4mm]
{\displaystyle\hat C_{kj}=(A_k-A_j)\hat S_z+\frac{1}{8\omega_I}(B_{k+}B_{k-}-B_{j+}B_{j-})\hat 1,}\\[4mm]
{\displaystyle\hat D_k=\lambda\hat 1+\sum_{s\not=k}A_s\hat I_{sz}+\frac{1}{8\omega_I}\left(4\omega_1^2-B_{k+}B_{k-}\right)\hat 1.}
\end{array}
\label{XY}
\end{equation}
The advantage of formulas (\ref{fa}) is that they reduce the inversion ${\bf L_{0,22}}^{-1}$ in the subspace ${\cal L}_C$ to the much simpler problem of inversions $\hat C'_{kj},\,\hat D'_k$ of Zeeman operators. The latter are defined in the $2^{n+1}$-dimensional Hilbert space and are diagonal in the usual Zeeman basis.

The last terms in the expressions for $\hat C_{kj}$, $\hat D_k$ are the second order corrections with respect to the inverse nuclear Larmor frequency $\omega_I^{-1}$. For typical random spin geometries, these terms are either quenched by the first order terms or small simultaneously with them. Hence, these terms can often be neglected.

\subsection{The Lindblad form}

Introducing the notation
$$
{\bf D}(\hat X)\hat\rho\equiv\hat X\hat\rho\hat X^\dagger-\frac{1}{2}\left\{\hat\rho,\hat X^\dagger\hat X\right\}
$$
and utilizing the double-commutator character of approximation (\ref{fa}) as well as the thermal correction ${\bf\Gamma_{th}}$, it is straightforward to see that the right-hand side of the Zeeman projected equation (\ref{aaZ}) is written in the purely Lindblad form
$$
{\bf L_Z}=\Gamma_+^{(S)}{\bf D}(\hat S_+)+\Gamma_-^{(S)}{\bf D}(\hat S_-)+
\sum_k\left(\Gamma_{k+}^{(I)}{\bf D}(\hat I_{k+})+\Gamma_{k-}^{(I)}{\bf D}(\hat I_{k-})\right)+
$$
$$
+\sum_k\hat\Gamma_k^{(IS)}{\bf D}(\hat Y_k)+\sum_{k<j}\hat\Gamma_{kj}^{(II)}{\bf D}(\hat X_{kj})
$$
with the constant rates
$$
\Gamma_\pm^{(S)}=\frac{1\mp P_0}{2}R_1^{(S)}+\frac{\omega_1^2}{2\omega_I^2}R_2^{(S)},
$$
$$
\Gamma_{k\pm}^{(I)}=\frac{1}{2}\left(R_1^{(I)}+\frac{\vert B_k\vert^2}{4\omega_I^2}R_2^{(I)}\right)
$$
related to the single-spin jump operators $\hat S_\pm$, $\hat I_{k\pm}$, and the operator rates
$$
\hat\Gamma_k^{(IS)}=\frac{\omega_1^2B_{k+}B_{k-}}{8\omega_I^2(R_2^{(S)}+R_2^{(I)})}(1+\hat D_k^2)^{-1},
$$
$$
\hat\Gamma_{kj}^{(II)}=\frac{d_{kj}^2}{4R_2^{(I)}}(1+\hat C_{kj}^2)^{-1},
$$
$$
\hat D_k=\frac{\lambda\hat 1+\sum_{s\not=k}A_s\hat I_{sz}}{R_2^{(S)}+R_2^{(I)}},\quad
\hat C_{kj}=\frac{(A_k-A_j)\hat S_z}{2R_2^{(I)}}
$$
related to the two-spin jump operators
$$
\hat Y_k=\hat I_{k+}\hat S_-+\hat I_{k-}\hat S_+,\quad
\hat X_{kj}=\hat I_{k+}\hat I_{j-}+\hat I_{k-}\hat I_{j+}.
$$

\subsection{Comparison between full Liouville space simulations and kMC simulations}
\begin{figure}[ht]
\includegraphics[width=0.7 \textwidth]{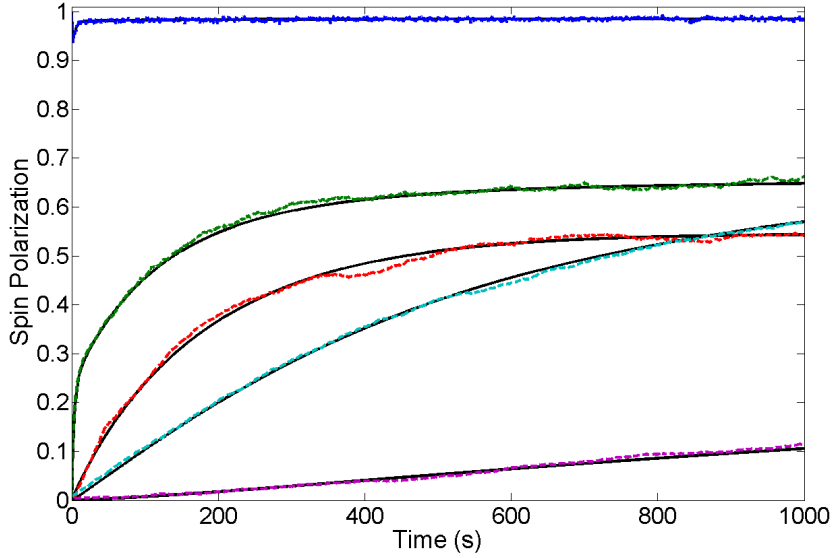}
\caption{\label{fig3} Comparison between a simulation of the polarization dynamics during SE DNP using the quantum master equation in the full Liouville space and a simulation based on a master equation for states contained in the Zeeman subspace that was obtained by adiabatic elimination and solved by kMC. For the kMC simulation $10^4$ trajectories were averaged. The polarization dynamics is calculated for a system containing one electron (blue) and four coupled nulcei. There is excellent agreement between both simulations with the coloured lines indicating the results obtained using kMC and the black lines showing the results using the quantum master equation. The agreement can be further improved by increasing the number of averaged trajectories. }
\end{figure}

We have carried out a number of comparisons between simulations of the polarization dynamics using the quantum master equation and our proposed method that involves an adiabatic approximation procedure to obtain a master equation for the states contained in the Zeeman subspace ${\cal L}_Z$ and the use of a kinetic Monte Carlo (kMC) algorithm to calculate the polarization dynamics. As long as the parameters are chosen in accordance with the conditions described in the supplementary material, an excellent agreement between the two simulation methods is achieved. An example of such a comparison is given in Fig. (\ref{fig3}) which shows the polarization dynamics in a system containing one electron and four nuclear spins forming a coupled network. The agreement can be further improved by increasing the number of averaged trajectories for the kMC method, however this is done on the expense of further increasing the computational time. The parameters used for the simulation are summarised in the following table.
\begin{table}[!h]
\begin{tabular}{|c |c|}
\hline \textbf{Parameter} & \textbf{Value}  \\ \hline
Nuclei & $^1H$ \\ \hline
B$_0$ & 3.4 T \\ \hline
Temperature & 1 K \\ \hline
Number of trajectories  & $10^4$ \\ \hline
$\omega_1$ & 50 kHZ \\ \hline
nuclear X-coord. & 3.51, -2.52, 5.98 -4.01  (\AA) \\ \hline
nuclear Y-coord. & 0.02, 0.03, 0.00 0.00  (\AA) \\ \hline
nuclear Z-coord. & 3.56, -4.34, -0.52 0.02  (\AA) \\ \hline
electron coord. & -0.03, -0.01, -0.01 (\AA) \\ \hline
$\vert B_+\vert$ & 0.935 MHz, 0.822 MHz, 92.1 kHz, 22.2 kHz \\ \hline
A & 0.318 MHz, 0.794 MHz, -0.352 MHz, -1.26 MHz \\ \hline
$d_{1,2}, d_{1,3}, d_{1,4}$ & -0.055 kHz, -0.660 kHz, 0.048 kHz \\ \hline
$d_{2,3}, d_{2,4}, d_{3,4}$ & 0.037 kHz, -1.040 kHz, 0.059 kHz \\ \hline
$T_1e$, $T_2e$, $t_1n$, $t_2n$ & 1 s, 10 $\mu$s, 1 h, 5 ms \\ \hline
\end{tabular}
\end{table}
 
\subsection{SE DNP and Spin Diffusion}

\begin{figure}[ht]
\includegraphics[width=0.7 \textwidth]{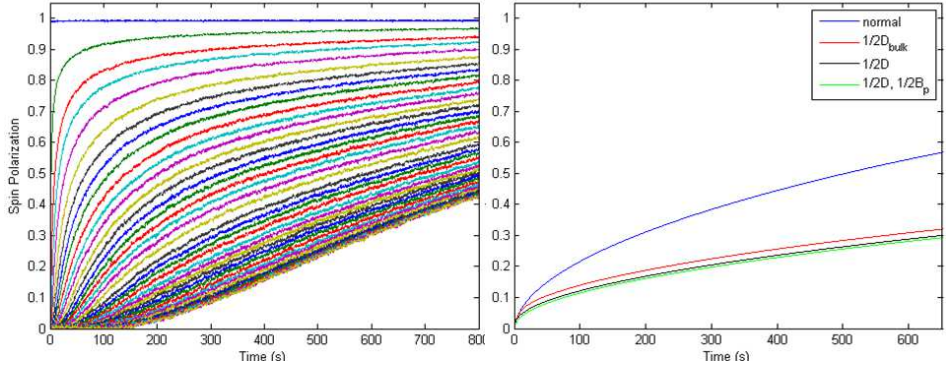}
\caption{\label{fig4} Polarization buildup in a linear chain of 40 coupled nuclear spins with an electron located at one end. The buildup curve for each individual nucleus is shown. On the right hand we show the average polarization for the spin chain for the cases that i) all dipolar constants are given by the spacing between the individual nuclei, ii) the dipolar coupling constants $d_{kj}$ are halved for bulk nuclei, iii) the dipolar coupling constants are halved for all nuclei and iv) the dipolar couplings are halved for all nuclei and in addition the pseudo-secular interaction strength between the electron and the first nuclei is reduced by $\sqrt{2}$}
\end{figure}

To evaluate the role that the nuclear dipolar interaction plays in the distribution of the polarization within a larger ensemble of  coupled nuclear spins, we have simulated the polarization buildup dynamics in a linear chain of 40 coupled spins with an electron located at one end. The individual buildup curves for the 40 nuclear spins are shown in Fig. (\ref{fig4}). The simulation parameters are listed in the following table (they are identical with the parameters used in \cite{hovav10}) .
\begin{table}[!h]
\begin{tabular}{|c |c|}
\hline \textbf{Parameter} & \textbf{Value}  \\ \hline
Nuclei & $^{13}C$ \\ \hline
$\omega_I$ & 36 MHz \\ \hline
Temperature & 1 K \\ \hline
Number of trajectories  & $10^5$ \\ \hline
$\omega_1$ & 100 kHz \\ \hline
$\vert B_+\vert_{e-n_1}$ & 40 kHz \\ \hline
$\vert B_+\vert_{e-n_{j>1}}$ & 0 kHz \\ \hline
A & 0 Hz \\ \hline
$\left<d\right>$ & 7.45 Hz $\pm$ 0.82 Hz \\ \hline
$T_1e$, $T_2e$, $t_1n$, $t_2n$ & 10 ms, 10 $\mu$s, $10^5$ s, 1.25 ms \\ \hline
\end{tabular}
\end{table}
To investigate the role of the nuclear dipolar coefficient $d_{kj}$ for the distribution of the polarization within the linear spin chain, we calculated the average polarization buildup and compared it to the buildup obtained when the diplar coefficient is set to half its value for all nuclei $k>1$ (fig. (\ref{fig4})). In further simulations we either set all dipolar coefficients to half their value or ,in addition, we decreased the pseudo-secular interaction by a factor $\sqrt{2}$. These changes of the interaction parameters are in analogy to the set of simulations carried out by Hovav \emph{et al.} in Ref. \cite{hovav10}. In contrast to their observations, the biggest effect on the average polarization is already brought about by halving the dipolar interaction strength for all nuclei $k>1$, indicating the importance of this interaction for the diffusive transport. Any additional changes only introduced further small decreases of the average polarization.

\begin{figure}[ht]
\includegraphics[width=0.48 \textwidth]{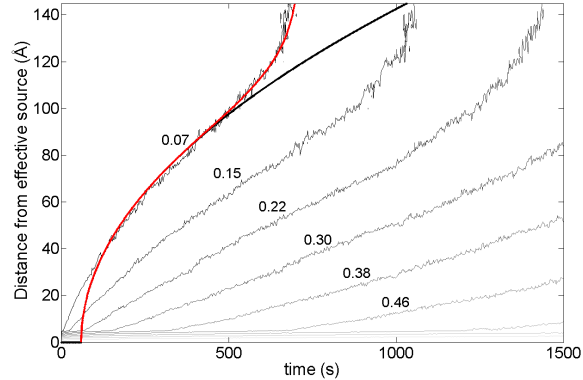}
\caption{\label{fig5} Comparison between a simple diffusion model and the results obtain by kMC simulations. The black contour lines visualise how nuclear spin polarization at various levels moves in time through a linear nuclear spin chain consisting of 30 dipolar coupled spins during a SE DNP. These results are compared to a simple diffusion model that is based on a polarization source at the start of the spin chain and an average diffusion constant as described in eq. (\ref{dif}). The solid black line shows how polarization would move through the linear chain assuming a fully absorbing boundary at the end of the chain and the red line indicates the time course of the polarization if a reflective boundary condition is assumed.}
\end{figure}

We then compared  a simple diffusion model $ \partial p/\partial t =  D_{\rm av} \frac{\partial^2 p}{\partial x^2}$, where $p\equiv p(x,t)$ is the polarization field and a constant source is assumed for the start of the linear spin chain. The average diffusion constant was obtained from eq. (\ref{dif}). This diffusion model was solved for two different boundary conditions, either with an absorbing boundary condition or a reflective boundary condition. The time course of the polarization moving through the linear spin chain is compared in Fig. (\ref{fig5}) with a contour plot derived from simulating the polarizationchanges within the linear spin chain using our method described in the main part of the paper. It is clear evident that the dynamics is reasonably well approximated by the simple diffusion model if a reflective boundary condition is used. The parameters for this simulation are provided in the following table.

\begin{table}[!h]
\begin{tabular}{|c |c|}
\hline \textbf{Parameter} & \textbf{Value}  \\ \hline
Nuclei & $^{13}C$ \\ \hline
B$_0$ & 3.4 T \\ \hline
Temperature & 1 K \\ \hline
Number of trajectories  & $10^5$ \\ \hline
$\omega_1$ & 50 kHz \\ \hline
separation & 5 $\AA$ $\pm$ 5\% \\ \hline
angle $\theta$ of chain to B$_0$ & 45$^\circ$ \\ \hline
$T_1e$, $T_2e$, $t_1n$, $t_2n$ & 1 s, 10 $\mu$s, $10^8$ s, 0.1 ms \\ \hline
\end{tabular}
\end{table}

\end{document}